\documentclass[letterpaper,aps,prl,twocolumn,amsmath,showpacs,amssymb]{revtex4-1}
\pdfoutput=1
\usepackage{color}
\usepackage[pdftex]{graphicx,hyperref}
\usepackage{dcolumn}
\usepackage{bm}
\usepackage{marvosym}
\usepackage{times,amsmath,amssymb}

\begin{document}

\title{Graphene-based polaritonic crystal}

\author{Yu. V. Bludov, N. M. R. Peres, M. I. Vasilevskiy}

\affiliation{Centro de F\'{\i}sica e Departamento de F\'{\i}sica, Universidade do Minho,
Campus de Gualtar, Braga 4710-057, Portugal
}
\begin{abstract} It is shown that monolayer graphene deposited on a spatially-periodic gate behaves as a polaritonic crystal.
Its band structure depending on the applied gate voltage is studied. The scattering of electromagnetic radiation from such a crystal is presented calculated and analyzed in terms of Fano-type resonances between the reflected continuum and plasmon-polariton modes forming narrow bands.

\end{abstract}

\pacs{81.05.ue,72.80.Vp,78.67.Wj}

\maketitle

Coupling of light to the surface charges at a
metal-dielectric interface gives rise to a
special kind of evanescent electromagnetic (EM) waves called surface
 plasmon-polaritons (SPPs) \cite{c:SPP-common}. The
specific properties of SPPs allow for their use in variety of
practical applications. The sensitivity of SPPs to the
properties of the dielectric, the metal and the interface is used in SPP-based
sensors \cite{c:sensors,Shalabney2011} and in high-resolution imaging \cite{Barnes2003,c:zayats,c:imaging}.
Surface plasmons give rise to very large EM fields at the surface, which is important for surface-enhanced optical spectroscopies \cite{Hartland2011}.
Moreover, the SPP wavelength can be much smaller than the photon wavelength, opening the
possibility for further miniaturization of photonics components,
a new field of research called {\it nanoplasmonics} \cite{c:plasmonics}.
Of particular interest is the ability to tune SPP modes in plasmonic devices by
external control: using an electric field in a liquid crystal \cite{c:liq-crys},
a magnetic field in a magneto-optically active substrate \cite{c:mag-f},
 thermal heating \cite{c:heating}, or a light beam focused on a non-linear coating \cite{c:nonlinear}.

The possibility of tuning the amount of free carriers in graphene using an external gate allows for an effective control of the
material's optical properties \cite{c:graphene-review,c:graphene-gating,c:graphene-optics,Koppens2011}.
Exploring graphene, a tunable two-dimensional (2D) metal, for plasmonics at the nano-scale reveals new physical effects and opens exciting possibilities in this field \cite{c:graphene-plasmons,c:graphene-plasmons-TE,c:disp-rel,c:graphene-plas-tuning}.
When compared to their counterparts in conventional 2D electron systems, SPPs in graphene exhibit some new and unusual properties, such as the $1/4$-power
density dependence of the SPP frequency \cite{c:graphene-plasmons}
and the existence of $s$-polarized waves \cite{c:graphene-plasmons-TE}.
Moreover, SPPs in graphene can potentially be used in a variety
of practical applications. For example, using the amplification of
SPPs in graphene opens the possibility to create a terahertz radiation
source \cite{c:graphene-plas-source}; employing the attenuated total reflection
(ATR) configuration with a gated graphene layer allows for a resonant switching of the reflection coefficient of an external EM wave from nearly unity to almost zero \cite{c:switch}.

\begin{figure}
\begin{center}
\includegraphics{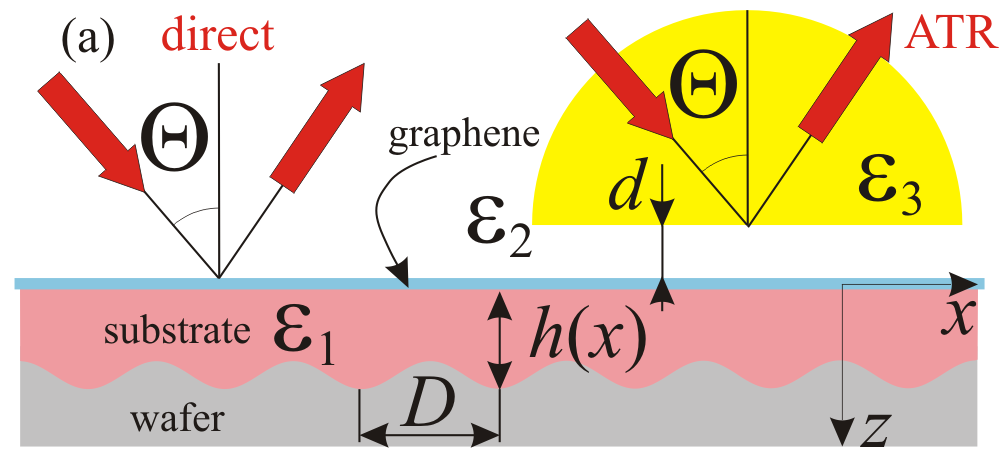}
\includegraphics[width=8.5cm]{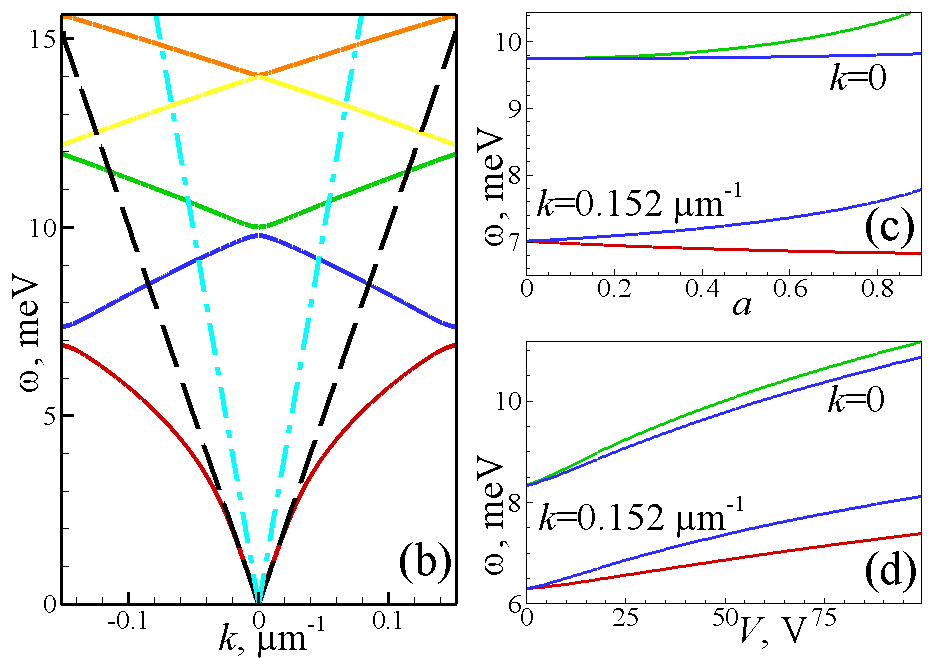}
\end{center}
\caption{ (color on line) (a) Geometry of the system:
"sandwich"-like structure containing a graphene layer on
a substrate composed of a dielectric spacer (with dielectric
constant $\epsilon _1$) and a transparent conductive wafer with
a periodic interface relief; (b)
Real part of the SPP frequency versus wave vector in the first
Brillouin zone, calculated for a single graphene layer at interface
of two semi-infinite dielectric media; (c, d) Edges of the first two
lowest gaps (at edge  and at center of Brillouin zone) versus
modulation depth $a$ (panel c) or gate voltage $V$ (panel d).
The parameters are the following: $\varepsilon_2=1$,
$\varepsilon_1=3.9$, $h_0=300\,$nm, $D=20.67\,\mu$m,
$a=0.6$ (panels b and d). In panels (b) and (c) $V=50$V, corresponding to an average chemical potential $\bar \mu\approx 0.222\,$eV. In panel (b) the light lines for the cladding media, $k c/\omega =\sqrt{\varepsilon_{1,2}}$, are shown dashed (1) and dash-dotted (2).
} \label{fig:eigen-val}
\end{figure}

Recently, a new class of metamaterials based on SPPs in
 graphene, either deposited on a periodically corrugated surface
\cite{c:waveguide} or composed of an array of micro-ribbons
\cite{Ju2011} was proposed, with a high potential interest for
 transformation optics. When SPPs propagate along a periodically
 modulated surface, the concept of a "surface polaritonic crystal"
can be introduced \cite{c:zayats}, where the SPP dispersion shows a
 band-gap structure \cite{c:band-gap-spp}, in analogy with a photonic
crystal. In this Letter we propose and theoretically analyze a new type
 of SPP crystal, based on a graphene sheet deposited on top of a periodically
 modulated gate electrode (wafer), as schematically represented in Fig. \ref{fig:eigen-val}(a).

In order to achieve the periodic modulation
of graphene's conductivity, we consider a single
graphene layer in the plane $z=0$ deposited on a SiO$_2$
 substrate with a dielectric constant
$\varepsilon_1$ [see Fig. \ref{fig:eigen-val}(a)].
The opposite side of the substrate has a periodic relief with
a spatial period $D$, such that $h(x)=h(x+D)$. To be specific, we will
consider the dielectric thickness modulation of the form
$h(x)=h_0[1+a\cos(gx)]$, where $a$ is the modulation depth and $g=2\pi/D$.
A conductive wafer is placed over this modulated surface, serving as a gate contact.
If a constant gate voltage $V$ is applied between the graphene layer
and the wafer, a periodic modulation of the garphene conductivity can be achieved.
If $D\gg h(x)$, the carrier density
in graphene can be expressed as $n(x)=\xi \varepsilon_1 V/4\pi e h(x)$,
where $e$ is the electron charge and $\xi $ is a coefficient between 1
and 2 depending on the charge distribution on the corrugated gate electrode (further in the Letter we use $\xi=1$).
Owing to the periodicity of the surface relief, the carrier density
in graphene is a periodic function with the same period $D$, thus
resulting in the periodicity of its optical conductivity,
$\sigma(x,\omega)=\sigma(x+D,\omega)$. The latter
can be related to the local value of the chemical potential counted with respect to the Dirac point, $\mu(x)=\hbar v_F \{\pi n(x)\}^{1/2}$,
where $v_F$ is the Fermi velocity.

As it is shown below, the periodic modulation of the optical
conductivity leads to the possibility of direct coupling of a
propagating EM wave to the surface plasmons (Fig. \ref{fig:eigen-val}(a),
left, where $\varepsilon_2$ is the dielectric constant of the medium above graphene).
However, only those SPP modes which lie within the
 light cone, $\omega /k=c/\sqrt {\varepsilon_2}$, shown by dash-dotted lines
in Fig. \ref{fig:eigen-val}(b), can be excited this way.
In general, for the SPP excitation one has to consider an ATR
structure like the one described in Ref. \onlinecite {c:switch},
which includes a prism with a dielectric constant $\varepsilon_3$
(Fig. \ref{fig:eigen-val}(a), right). Usually there is a
 gap between the graphene sheet and the prism, which we shall model as a dielectric layer of a thickness $d$ and a dielectric constant $\varepsilon_2$. We assume that the prism occupies the half-space $z<-d$ and a $p$-polarized EM wave impinges on the boundary $z=-d$, coming from $z=-\infty$ at an angle of incidence $\Theta$.

Since the dielectric properties of the structure are periodic along $x$, the solution of Maxwell's equations,
${\rm rot}{\textbf{E}^{(m)}}=i\kappa\textbf{H}^{(m)}$,
${\rm rot}{\textbf{H}^{(m)}}=-i\kappa\varepsilon_m\textbf{E}^{(m)}$
for the $p$-polarized wave [with components $\bm E=(E_x,0,E_z)$
 and $\bm H=(0,H_y,0)$] can be written as Fourier-Floquet series:
\begin{eqnarray}
H^{(m)}_y(x,z)&=&\sum_{n=-\infty}^\infty\left[A_{n}^{(m)}\exp(\kappa q^{(m)}_n z)+\right.\label{eq:sol1}\\
&&\left.B_{n}^{(m)}\exp(-\kappa q^{(m)}_n z)\right]\exp([i(k+ng)x], \nonumber\\
E^{(m)}_x(x,z)&=&\sum_{n=-\infty}^\infty\frac{q^{(m)}_n}{i\varepsilon_m}\left[A_n^{(m)}\exp(\kappa q^{(m)}_n z)-\right.\label{eq:sol2}\\
&&\left.B_n^{(m)}\exp(-\kappa q^{(m)}_n z)\right]\exp([i(k+ng)x]\nonumber.
\end{eqnarray}
In Eqs. (\ref{eq:sol1}) and (\ref{eq:sol2}), $q^{(m)}_n=\sqrt{\left[(k+ng)/\kappa \right]^2-\epsilon_m}$ is the in-plane component of the photon
wavevector in the medium $m$ ($m=1,2,3$; $m=3$ applies only for the ATR configuration), $\kappa=\omega/c$, and $c$ is the speed of light in vacuum.
Although the substrate is finite, for simplicity we shall consider the medium 1 as semi-infinite in the wave equations.
If the gate electrode is transparent, this simplification is not crucial for the analysis of the optical properties of the modulated structure.
Boundary conditions at $z=-d$ imply the continuity of the tangential components of the electric and
magnetic fields,  [$E_x^{(3)}(x,-d)=E_x^{(2)}(x,-d)$, $H_y^{(3)}(x,-d)=H_y^{(2)}(x,-d)$].
At $z=0$, the tangential component of the electric field is
 continuous, $E^{(1)}_x(x,0)=E^{(2)}_x(x,0)$, while
 the discontinuity of the tangential component of
the magnetic field,  $H^{(1)}_y(x,0)-H^{(2)}_y(x,0)
=-(4\pi/c)j_x=-(4\pi/c)\sigma(x,\omega)E_x(x,0)$,
stems from the presence of surface currents (caused by the SPP electric field) in the graphene layer.
Applying these boundary conditions, one can find the
explicit form of the transfer matrices, $\hat{M}^{m\leftarrow m+1}$, which relate the coefficients in Eqs. (\ref{eq:sol1}) and (\ref{eq:sol2}) for different $m$,
 $(\ldots, A_{n}^{(m)},B_{n}^{(m)},A_{n+1}^{(m)},B_{n+1}^{(m)},\ldots)^T=\hat{M}^{m\leftarrow m+1}(\ldots, A_{n}^{(m+1)},B_{n}^{(m+1)},A_{n+1}^{(m+1)},B_{n+1}^{(m+1)},\ldots)^T$. The matrices $\hat{M}^{m\leftarrow m+1}$ consist of 2$\times$2 blocks $\hat{M}^{m\leftarrow m+1}_{n,l}$,
\begin{eqnarray}
\hat{M}^{1\leftarrow 2}_{n,l}=\frac 12 \left(
\begin{array}{cc}
Q_{n,l}^{(1,+)}-\frac{4\pi q_n^{(2)}}{ic\varepsilon_2}\sigma_{n-l} & Q_{n,l}^{(1,-)}+\frac{4\pi q_n^{(2)}}{ic\varepsilon_2}\sigma_{n-l} \\
Q_{n,l}^{(1,-)}-\frac{4\pi q_n^{(2)}}{ic\varepsilon_2}\sigma_{n-l} & Q_{n,l}^{(1,+)}+\frac{4\pi q_n^{(2)}}{ic\varepsilon_2}\sigma_{n-l}
\end{array}\right)\nonumber\\
\hat{M}^{2\leftarrow 3}_{n,l}=\frac 12 \left(
\begin{array}{cc}
Q_{n,l}^{(2,+)}e^{\kappa\left[q_l^{(2)}-q_l^{(3)}\right]d} & Q_{n,l}^{(2,-)}e^{\kappa\left[q_l^{(2)}+q_l^{(3)}\right]d} \\
Q_{n,l}^{(2,-)}e^{-\kappa\left[q_l^{(2)}+q_l^{(3)}\right]d} & Q_{n,l}^{(2,+)}e^{-\kappa\left[q_l^{(2)}-q_l^{(3)}\right]d}
\end{array}\right)\nonumber
\end{eqnarray}
where
$Q_{n,l}^{(m,\pm)}=\delta_{n,l}\left(1\pm F_n^{(m)}\right)$, $F_n^{(m)}=
\varepsilon_mq_n^{(m+1)}/\varepsilon_{m+1}q_n^{(m)}$, $\delta_{n,l}$
is a Kronecker symbol, $\sigma_{n}=D^{-1}\int_0^{D}{\sigma(x,\omega)}\exp(-ingx)dx$ is
the spatial Fourier transform of the graphene conductivity.

\begin{figure}
\begin{center}
\includegraphics[width=8.5cm]{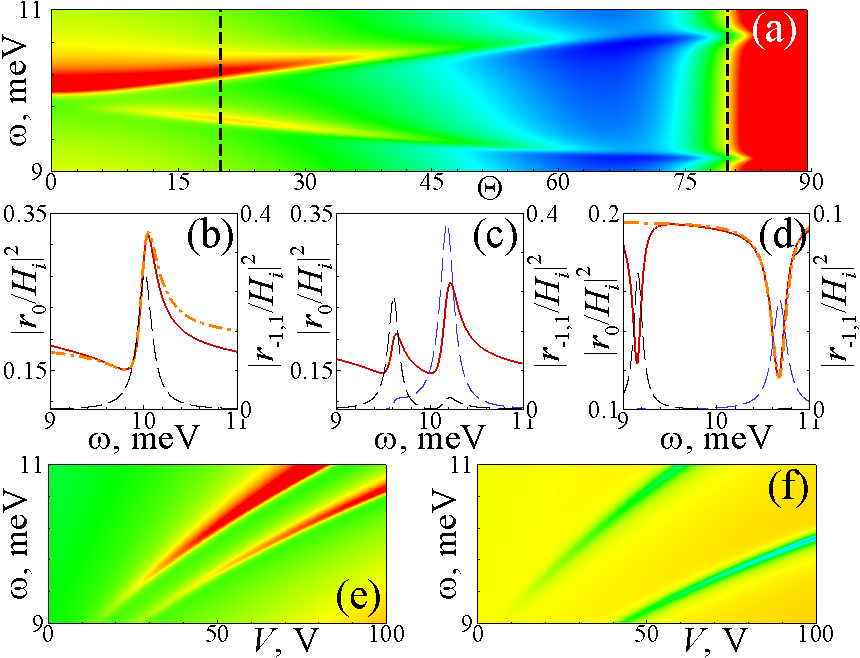}
\end{center}
\caption{ (color on line) (a) Zero harmonic reflection coefficient, $|r_0/H_i|^2$, \textit{vs} frequency and incidence angle for SPPs in periodically modulated graphene with $V=50\,$V; (b--d) Zero harmonic reflection coefficient (red solid lines) and relative square amplitudes, $|r_n/H_i|^2$, of $n=1$ (blue dashed lines) and $n=-1$ (black dashed lines) harmonics for the incidence angles $\Theta=0^o$ (b), $\Theta=20^o$ (c), and  $\Theta=80^o$ (d) [the latter two correspond to the vertical dashed lines in panel (a)]; (e, f) Zero harmonic reflection coefficient \textit{vs} frequency and gate voltage for $\Theta=20^o$ (e), and  $\Theta=80^o$ (f). Other parameters are the same as in Fig. \ref{fig:eigen-val}. In panels (a, e, f) red (blue) color corresponds to high (low) values of the reflection coefficient. In panels (b, d) Fano-type fits are shown by dash-dotted (orange) lines, with $R_0=0.1508$, $\delta R=0.035$, $q=2$, $\omega_0=10 $meV (b) and $R_0=0.116$, $\delta R=0.08$, $q=0.08$, $\omega_0=10.68 $meV (d); $\gamma=0.1 $meV.
} \label{fig:exitation}
\end{figure}

The meaning of the coefficients $A_n^{(m)}$, $B_n^{(m)}$ is different for the different media.
The waves corresponding to the different terms in Eqs. (\ref{eq:sol1}) and (\ref{eq:sol2}) can be either propagating (with
${\rm Im}(q^{(m)}_n)<0$) or evanescent (with ${\rm Re}(q^{(m)}_n)>0$). Since the incident wave in the medium 3 possesses only the $n=0$ component (zero harmonic), ${\rm Im}(q^{(3)}_0)<0$
and the coefficients $B_n^{(3)}=\delta_{n,0}H_i\exp(-\kappa q_0^{(3)}d)$ are
proportional to the magnetic field amplitude ($H_i$) in the incident wave.
In the medium 1, the coefficients $A_n^{(1)}\equiv 0$ correspond to the
absence of the corresponding harmonics coming from $z=\infty$. Then,
multiplying the matrices $\hat{M}^{1\leftarrow 2}\hat{M}^{2\leftarrow 3}$,
and taking into account the block-diagonal structure of the matrix $\hat{M}^{2\leftarrow 3}$, after some algebra we obtain the following equations for the amplitudes of the reflected harmonics, $r_n=A_n^{(3)}\exp(-\kappa q^{(3)}_nd)$:
\begin{equation}
\hat{R}\times (\ldots, r_n, r_{n+1}, \ldots)^T=(\ldots,
H^{i}_{n^\prime}, H^{i}_{n^\prime+1}, \ldots)^T\,,
\label{eq:ref-ampl}
\end{equation}
where the elements of the matrix $\hat{R}$ and the vector $H^{i}_n$ are
\begin{eqnarray}
R_{n,n^\prime}=\left[\delta_{n,n^\prime}F_n^{(1)}-\frac{4\pi q_{n^\prime}^{(2)}\sigma_{n-{n^\prime}}}{ic\varepsilon_2}\right]\left[S_{n^\prime}+F_{n^\prime}^{(2)}C_{n^\prime}\right]+\nonumber\\
\delta_{n,{n^\prime}}\left[C_{n^\prime}+F_{n^\prime}^{(2)}S_{n^\prime}\right],\\
H^{i}_n=-\left\{\left[\delta_{n,0}F_0^{(1)}-\frac{4\pi q_0^{(2)}\sigma_{n}}{ic\varepsilon_2}\right]\left[S_0-F_0^{(2)}C_0\right]+\right.\nonumber\\
\left.\delta_{n,0}\left[C_0+F_0^{(2)}S_0\right]\right\}H_i,
\end{eqnarray}
with $C_n=\cosh(\kappa q^{(2)}_nd)$ and $S_n=\sinh(\kappa q^{(2)}_nd)$.

In order to obtain the general properties of SPPs in graphene with periodically modulated conductivity,
 we first consider the eigenvalue problem for the matrix $\hat{R}$ leading to
the dispersion relation for SPPs in a flat 2D graphene layer placed between two lossless dielectric media ($\varepsilon_1$
and $\varepsilon_2$). We put $\varepsilon_3=\varepsilon_2$ and solve the equation ${\rm det}(\hat{R})=0$.
It yields complex eigenvalues because the graphene conductivity $\sigma(x,\omega)$ has both real and imaginary parts, therefore the SPP eigenmodes are dissipative.
The SPP dispersion curve for the real part of the frequency eigenvalue, $\omega^\prime$, \textit{vs} wavenumber for the first Brillouin
zone, $k\in[-g/2,g/2]$, is presented in Fig. \ref{fig:eigen-val}(b). The imaginary part of the frequency (mode damping) is an order of magnitude smaller than $\omega^\prime$.
As it can be seen from Fig. \ref{fig:eigen-val}(b),
the SPP dispersion curve is periodic in the $k$-space, with the period $g$. There are bands of allowed SPP frequencies, separated by gaps opening at the edges and
in the center of the Brillouin zone, where SPPs do not exist.
As expected, the widths of the gaps increase with the
increase of the modulation depth $a$ [Fig.\ref{fig:eigen-val}(c)].
A natural question arises: \textit{is it possible to control
dynamically the gap widths through some nondestructive external knob?}
The positive answer to this question is evident from Fig. \ref{fig:eigen-val}(d). Since the chemical potential of
graphene can be tuned by the gate voltage, one can shift the
spectral position and width of the gaps by changing $V$.
Therefore, the SPP crystal band structure can be controlled dynamically.

Another feature of the SPP spectrum in periodically
modulated graphene is that the "scan line", $k=\kappa \sqrt {\varepsilon_2}\sin \Theta\:,$ located within the dash-dotted lines in Fig. \ref{fig:eigen-val}(b)
crosses the SPP dispersion curves. This situation is completely
different from the case of uniform graphene \cite{c:switch},
where both phase and group velocities of SPPs are smaller
than the velocities of light in the surrounding dielectrics.
SPPs in periodically modulated graphene can be excited by an
external propagating EM wave, without an ATR prism.
This is illustrated by Fig. \ref{fig:exitation}, where the
amplitudes of the reflected field harmonics are presented; they have been
calculated by solving Eq. (\ref{eq:ref-ampl})
for $d=0$ and $\varepsilon_3=\varepsilon_2$ (in this case, $\omega$ is real).

At normal incidence ($\Theta=0$), the zero harmonic
 reflection coefficient of the SPP crystal exhibits
just one maximum at $\omega\approx 10\,$meV
[see Figs. \ref{fig:exitation}(a) and (b)], which
approximately corresponds to the upper edge of the
second gap in Fig. \ref{fig:eigen-val}(b).
This is related to the parity of the SPP
mode with respect to $x=0$, since it is excited by a plane wave and $h(x)$ is an even function. Also from Fig. \ref{fig:exitation}(b) it
is clearly seen that the enhanced reflection of the zero
 harmonic corresponds to the excitation of the SPP
harmonics with $n=-1$ and $n=1$. They correspond to the
 bottom of the third allowed SPP band and, for normal
incidence, are mixed into the $k=0$ band bottom mode.
 For oblique incidence [Figs. \ref{fig:exitation}(c)-(f)],
 there are \textit{two} resonances corresponding to the
second and third SPP bands and producing reflected field
harmonics with $n=\pm 1$. SPPs are effectively excited
when the frequency and the in-plane component of the
wave vector of the incident EM wave match those
of SPP eigenmodes of the modulated graphene. The energy of the incident wave is transferred
 into the SPP harmonics with $n=\pm 1$, while the refleted wave in the far field contains only zero harmonic. Note that the position and the amplitude of the resonances can be controlled by changing the gate
voltage [Figs. \ref{fig:exitation}(e) and \ref{fig:exitation}(f)].
Quite interestingly, the direct excitation of $n=\pm 2$ and higher SPP bands can produce {\it propagating} EM waves with $|k\pm g |\le \kappa \sqrt {\varepsilon_2}$, scattered at angles
$$
\Theta _{\pm 1}=\arcsin {\left (\frac{k\pm g}{\kappa \sqrt {\varepsilon_2}} \right )}\,.
$$

Let us focus on the characteristic spectral dispersion of the zero harmonic reflection coefficient around each of the SPP resonance
frequencies [Figs. \ref{fig:exitation}(b) and (c)] where an intrinsic
mode of the SPP crystal is excited.
This is an asymmetric Fano-type resonance with a
maximum accompanied by a neighboring minimum. a
phenomenon first discovered in the ionization spectra
\cite{Fano} of He and now known in many different branches
of physics including light scattering by photonic crystals \cite{Fano-PhC}.
If a discrete mode couples to a continuum of excitations,
the Fano resonance arises from the constructive (destructive)
interference of the localized and delocalized waves, taking
 place above (below) the discrete mode frequency $\omega _0$.
A simple analytic expression to describe such a
lineshape was suggested by Fano \cite{Fano}:
\begin{equation}
F(\omega )= \frac {(q\gamma +\omega - \omega_0)^2}{(\omega - \omega_0)^2 + \gamma ^2}\:,
\label{Fano}
\end{equation}
where $\gamma $ is the mode damping and $q$ is an assymetry parameter related to the relative strengths of the transitions associated with the discrete and continuum modes.
Fig. \ref{fig:exitation}(b) shows a Fano-type fit, $R(\omega )=R_0 + \delta R\cdot F(\omega ),$ to the calculated reflectivity spectrum ($R_0$ and $\delta R$ are constants).
The coupling between the continuum of propagating EM modes in the medium 2 and the SPP Bragg mode of the polaritonic crystal decreases with the increase of the incidence angle. When $\Theta$ is close to the Brewster angle, $\Theta_b={\rm atan}(\sqrt{\varepsilon_1/\varepsilon_2})\approx 63^o$,
the resonant excitation of $n=\pm 1$ modes results in two narrow and almost symmetric minima in the spectrum [see Fig. \ref{fig:exitation}(d) where the Fano fit corresponds to $q=0.08$, compared to $q=2$ in Fig. \ref{fig:exitation}(b)].

\begin{figure}
\begin{center}
\includegraphics[width=8.5cm]{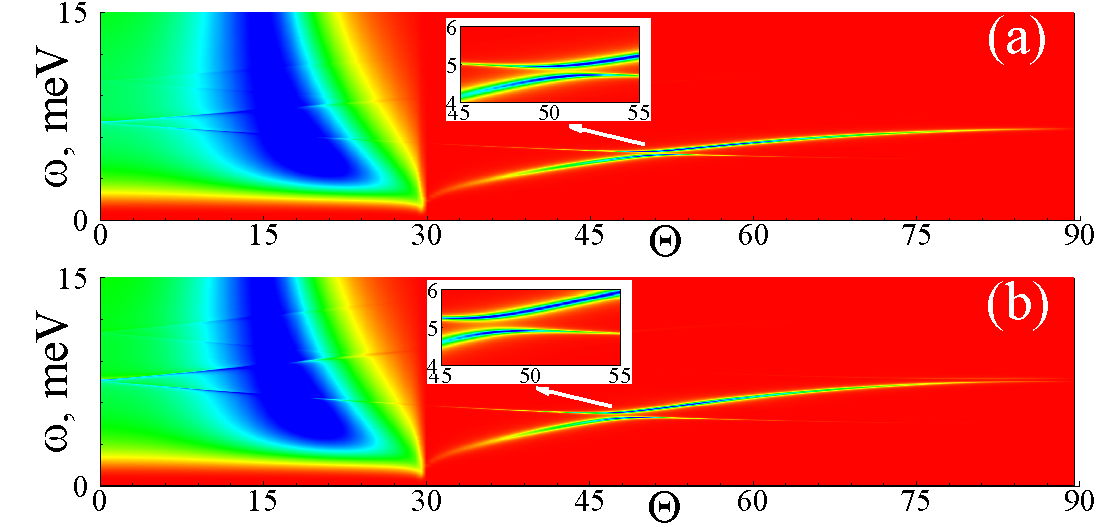}
\end{center}
\caption{ (color on line) Reflection
coefficient of zero harmonic $|r_0/H_i|^2$ \textit{vs}
frequency $\omega$ and angle of incidence $\Theta$ for
 ATR excitation of SPPs in graphene with parameters
$\varepsilon_3=16$, $\varepsilon_2=1$,
$\varepsilon_1=3.9$, $h_0=300\,$nm, $D=10.34\,\mu$m,
$a=0.6$, $d=10\,\mu$m, $V=50$V (panel a), or 90V (panel b).}
\label{fig:exitation-ATR}
\end{figure}

As mentioned above, the direct excitation of SPPs by a
propagating wave allows for probing only the part of the polaritonic crystal band structure comprised between the dash-dotted light lines in Fig. \ref{fig:eigen-val}(b). Beyond this, one has to use the ATR scheme where SPPs are excited by an evanescent wave with a sufficiently large wavenumber,
$
k=\kappa \sqrt {\varepsilon_3}\sin \Theta\:.
$
The results calculated for the
ATR structure are presented in Fig. \ref{fig:exitation-ATR},
where the SPP excitation conditions correspond to a minimum
of the zero harmonic reflectance related to the gap between the first and the second SPP bands. The mode
anticrossing, corresponding to the edges of the gap, is clearly seen in Figs. \ref{fig:exitation-ATR}(a)
and \ref{fig:exitation-ATR}(b) (near $\Theta\approx 50^o$,
 see insets). Comparison of Figs. \ref{fig:exitation-ATR}(a)
and \ref{fig:exitation-ATR}(b) shows, that increasing the gate voltage results in an increase of the gap width, as it could be anticipated from Fig. \ref{fig:eigen-val}(f).

To conclude, we have demonstrated that a single graphene layer deposited on a "sandwich"-like structure with a periodically corrugated gate electrode has the properties of a polaritonic crystal, namely, possesses a band
structure with gaps that can be tuned by the gate voltage. This crystal exhibits Fano-type resonances in the reflectance of incident EM waves due to the excitation of plasmon-polariton Bragg modes.

\end{document}